Contact Induced Spin Relaxation in Graphene Spin Valves with Al$_2$O$_3$ and MgO Tunnel Barriers

Walid Amamou[1], Zhisheng Lin[2], Jeremiah van Baren[2], Serol Turkyilmaz[3], Jing Shi[1,2], Roland K. Kawakami[1,2,4]*

[1]Program of Materials Science and Engineering, University of California, Riverside, CA 92521
[2]Department of Physics and Astronomy, University of California, Riverside, CA 92521
[3]Department of Electrical and Computer Engineering, University of California, Riverside, CA 92521
[4]Department of Physics, The Ohio State University, Columbus, OH 43210

*e-mail: kawakami.15@osu.edu

**Abstract**

We investigate spin relaxation in graphene by systematically comparing the roles of spin absorption, other contact-induced effects (e.g. fringe fields, etc.), and bulk spin relaxation for graphene spin valves with MgO barriers, Al$_2$O$_3$ barriers, and transparent contacts. We obtain effective spin lifetimes by fitting the Hanle spin precession data with two models that include or exclude the effect of spin absorption. Results indicate that additional contact-induced spin relaxation other than spin absorption dominates the contact effect. For tunneling contacts, we find reasonable agreement between the two models with median discrepancy of ~20% for MgO and ~10% for Al$_2$O$_3$.

I. Introduction

Spin relaxation is one of the most important issues in graphene spintronics. Theoretically, the spin lifetime of graphene is predicted to be on the order of microseconds due to weak spin-orbit coupling.[1-6] Experimentally, however, the measured spin lifetimes typically vary from about a hundred picoseconds to a few nanoseconds.[7-15] This discrepancy of three orders of magnitude represents one of the mysteries of the field. Understanding the origin of this discrepancy between theory and experiment will help close the gap and could lead to graphene materials and devices with exceptionally long spin lifetimes and spin diffusion lengths at room temperature.

Central to the investigation of spin relaxation is the role of contacts in graphene spin valves. Experimental studies are based on spin precession measurements in the nonlocal geometry[16] and modeling to extract the spin lifetime. In early studies, it was shown that high quality tunnel barriers between the ferromagnetic electrodes and graphene are needed to achieve spin lifetimes in excess of 200 ps,[13] but the nature of the contact-related effects producing this enhancement has remained unclear. One contact-related issue is the conductivity mismatch between the ferromagnetic electrode and the graphene, which causes the spins to flow from the graphene into the ferromagnet ("spin absorption") and results in lower effective spin lifetimes.[9,17] Other contact effects could include the nature of the interface between the ferromagnet and the graphene such as spin relaxation due to magnetostatic fringe field at the contact,[18] interfacial spin flip scattering[19-21] contact induced band bending, spin scattering due to the hybridization of the Co d-orbital with the graphene's π-orbital.[22-26] Therefore, the extracted spin lifetime will include contributions from the "bulk" spin relaxation in graphene (independent of contacts and including intrinsic and extrinsic spin scattering) and contact-induced spin relaxation. Because accurate determination of the bulk spin lifetime is important, it is worthwhile to clarify the role of contacts through systematic experimental studies.

In this Letter, we investigate spin relaxation in graphene by systematically comparing the roles of spin absorption, other contact-induced effects, and bulk spin relaxation. We analyze a set of graphene spin valves with Ti-seeded MgO barriers and utilize two different models of nonlocal Hanle precession to understand the role of spin absorption. Primarily, we find that some form of contact-induced spin relaxation other than spin absorption dominates the contact effect. To understand whether this is specific to MgO barriers or is more general, we develop graphene spin valves with smooth $Al_2O_3$ barriers and observe similar behavior. Beyond this, we are able to draw several conclusions, which can be described more clearly after discussing the models employed in the analysis.

To analyze Hanle spin precession data, we employ two different models. The first model is a "traditional model" (TM) which does not explicitly take spin absorption into account.[27] The spin accumulation in an applied magnetic field is described by the steady state Bloch equation:

$$D\nabla^2 \mu_s - \frac{\mu_s}{\tau_s} + \omega_L \times \mu_s = 0 \qquad (1)$$

where $D$ is the diffusion constant, $\tau_s$ is the bulk spin lifetime, $\mu_s$ is the spin accumulation, $\omega_L = g\mu_B B/\hbar$ is the Larmor frequency, $g$ is the gyromagnetic factor, $\mu_B$ is the Bohr magneton, and $B$ is the magnetic field. For the nonlocal Hanle geometry, this results in a nonlocal signal given by

$$R_{NL} \propto \pm \int_0^\infty \frac{1}{\sqrt{4\pi Dt}} \exp\left[-\frac{L^2}{4Dt}\right] \cos(\omega_L t) \exp(-t/\tau_s) \, dt \qquad (2)$$

where $L$ is the separation between spin injector and spin detector. Because the model does not explicitly take contact-induced spin relaxation into account, the fitted spin lifetime, which we denote as $\tau_{TM}$, should be considered as an effective spin lifetime given by,

$$\frac{1}{\tau_{TM}} = \Gamma_{bulk} + \Gamma_{abs} + \Gamma_{other} = \frac{1}{\tau_s} + \frac{1}{\tau_{abs}} + \frac{1}{\tau_{other}} \qquad (3)$$

where $\Gamma_{bulk}$ is the bulk spin relaxation rate in graphene, $\Gamma_{abs}$ is the contact-induced spin relaxation rate due to spin absorption, and $\Gamma_{other}$ is the contact-induced spin relaxation rate from other effects (e.g. fringe fields, etc.).

The second model explicitly takes into account the contact-induced spin relaxation by spin absorption, and we denote such models as "spin absorption models" (SAM).[9,17,28,29] This was first introduced by van Wees and coworkers[9,17], but an analytical solution for the Hanle curve was not found initially. Subsequently, Aji and coworkers[29] derived an analytical solution for the nonlocal Hanle signal, which is given by

$$R_{NL}^{\pm} = p_1 p_2 R_N f$$

$$f = Re \left( 2 \left[ \sqrt{1 + i\omega_L \tau_{SAM}} + \frac{\lambda}{2} \left( \frac{1}{r_1} + \frac{1}{r_2} \right) \right] e^{\left(\frac{L}{\lambda}\right)\sqrt{1+i\omega_L \tau_{SAM}}} + \frac{\lambda^2}{r_1 r_2} \frac{\sinh\left[\left(\frac{L}{\lambda}\right)\sqrt{1+i\omega_L \tau_{SAM}}\right]}{\sqrt{1+i\omega_L \tau_{SAM}}} \right)^{-1} \qquad (4)$$

where $\tau_{SAM}$ is the fitted spin lifetime, $\lambda$ is the spin diffusion length in graphene, and $r_i = \frac{R_F + R_C^i}{R_{SQ}} W$ with $i = 1$ for the injector and $i = 2$ for the detector, $R_C$ is the contact resistance, $R_{SQ}$ is the sheet resistance of graphene, $W$ is the width of the graphene, $R_N = R_{SQ}\lambda/W$ is the spin resistance of graphene, $R_F = \rho_{Co}\lambda_{Co}/A$ is the spin resistance of Co, $\rho_{Co}$ is the resistivity of Co, $\lambda_{Co}$ is the spin diffusion length in cobalt, and A is the junction area. Considering equation (2), the main difference from $\tau_{TM}$ is that the effect of spin absorption is explicitly calculated within the model, so the effective spin lifetime consists only of the remaining spin relaxation mechanisms, namely

$$\frac{1}{\tau_{SAM}} = \Gamma_{bulk} + \Gamma_{other} = \frac{1}{\tau_s} + \frac{1}{\tau_{other}} \qquad (5)$$

Thus, it is clear that the spin absorption model is more accurate than the traditional model in determining the bulk spin lifetime.

The large difference in spin resistances between graphene ($R_N$) and Co ($R_F$) will result in rapid flow of spins from graphene to Co. The amount of spin absorption can be controlled by introducing a tunnel barrier between the graphene and the ferromagnet. This is especially important for Co electrodes on graphene channel where the spin resistance of Co is much less than the spin resistance of graphene, which would produce large spin absorption in the absence of a barrier.

Beyond the van Wees and Aji model, the Otani group introduced an anisotropic term represented by the mixing conductance $G_{\uparrow\downarrow}$ that accounts for different spin absorption rate for longitudinal and transverse spins. While this anisotropy has been demonstrated in metallic spin valve, in the case of graphene the anisotropic spin absorption is almost the same as the isotropic one (within 4% difference) due to the higher junction resistance relative to the ferromagnet [28]. Therefore, in the case of graphene spin valves this model is equivalent to the van Wees and Aji model (equation 4).

Recently in Idzuchi et al [30], the Otani group re-analyzed our Hanle data on graphene spin valves with Ti-seeded MgO barriers using the spin absorption model. It was noticed that $\tau_{SAM}$ for four devices approached similar values near 500 ps, independent of the contact resistance-area product ($R_CA$). The interpretation was that the contact-induced effect is primarily due to spin absorption because the $\tau_{SAM}$ converges to a constant value representing the bulk spin lifetime (i.e. $\Gamma_{other} \sim 0$). Taken at face value, this is a good method to analyze spin relaxation: if $\tau_{SAM}$ converges to a constant value independent of $R_CA$, this likely represents the bulk spin lifetime; if $\tau_{SAM}$ varies and is correlated to $R_CA$, then the contact-induced spin relaxation from other effects ($\Gamma_{other}$) can be substantial; if $\tau_{SAM}$ varies and is uncorrelated to $R_CA$, other dependencies should be explored, such as the dependence on electrode spacing or other relevant parameters.

In our study, we follow this approach to analyze a larger set of graphene spin valves with Ti-seeded MgO barriers and $Al_2O_3$ barriers. We also compare the effective spin lifetimes from the two models, $\tau_{TM}$ and $\tau_{SAM}$, to elucidate the roles of spin absorption, other contact effects, and bulk spin relaxation in determining the overall spin lifetime. Through our studies, we draw the following conclusions: (1) the contact-induced spin relaxation rate from other contact effects, $\Gamma_{other}$, is substantial, (2) spin absorption is not the dominant source of spin relaxation, except in some cases with transparent contacts, (3) for tunneling contacts, the ratio $\tau_{TM}/\tau_{SAM}$ is typically found in the 60% - 90% range with median value of ~80% for MgO barriers and 70% - 100% range with median value of ~90% for $Al_2O_3$ barriers, (4) simulations provide a guideline for estimating the relative importance of spin absorption in determining the spin lifetime.

**II. Materials and Experimental Methods**

To fabricate devices, we exfoliate single layer graphene flakes onto $SiO_2$ (300 nm)/Si wafers, where the degenerately doped Si is used as a backgate. The graphene thickness is determined by optical contrast and calibrated using Raman spectroscopy. For the Ti-seeded MgO tunnel barrier devices, we utilize a single e-beam lithography pattern using bilayer PMMA/MMA resist combined with angle evaporations to define the MgO barrier and the 80 nm thick Co electrodes. The angle evaporation results in a small contact width of ~50 nm between the Co and graphene. Details of the device fabrication are provided elsewhere.[12] For the $Al_2O_3$ tunnel barrier devices, we grow aluminum over the entire sample by sputter deposition at a rate of 0.4 Å/s using Ar gas at a pressure of 5 mT and oxidize for 30 min at atmospheric pressure of $O_2$, following Dlubak *et* *al.*[31,32] The resulting $Al_2O_3$ layer can therefore act as both a tunnel barrier and a protective layer for the graphene. An AFM image of 1 nm of aluminum oxide on top of single layer graphene is shown in Figure 1(a). The rms roughness is 0.117 nm (within the yellow rectangle) and the surface is smooth with no observable pinholes at 1 nm thickness. This suggests that sputtered aluminum with post oxidation could be an excellent tunnel barrier candidate. To check for possible sputtering-induced defect formation, we perform Raman spectroscopy before and after $Al_2O_3$ deposition, shown in Figure 1(b). We observe the emergence of a relatively small D peak (30% of G peak magnitude), which suggests the presence of fewer sputtering-induced defects in the graphene layer compared to previous studies.[31,32] Finally, we define Co electrodes with various widths using bilayer PMMA/MMA resist and deposit 80 nm of Co in a molecular beam epitaxy (MBE) chamber with base pressure of $5\times10^{-10}$ torr. Because this process does not utilize the angle evaporation, the contact width is equal to the electrode width, which typically varies from 150 nm to 500 nm.

To characterize the electrical properties of the $Al_2O_3$ tunnel barrier, we measure gate dependence and three terminal differential contact resistance dV/dI. First, we perform a 4-probe resistance measurement as a function of gate voltage to determine the charge neutrality point and field effect mobility of the graphene. The measurement is performed using lock-in detection with an AC current of 1 µA at 211 Hz injected between contacts E1 and E4 (inset figure 1(c)) and voltage detection between E2 and E3. Figure 1(c) shows a typical gate dependence curve of graphene with a 1 nm $Al_2O_3$ overlayer. We observe a relatively small doping effect with charge neutrality point at $V_G$ = 15 V. The resulting carrier density at zero gate voltage is $1.08\times10^{12}$ $cm^{-2}$ (holes). The electron mobility is extracted from the slope of the gate dependent conductivity σ, and is $\mu_e$= 3005 cm$^2$/Vs for electrons and $\mu_h$= 948 cm$^2$/Vs for holes. The values of the mobility are comparable to pristine graphene spin valves [8,12].

We characterize the tunneling property of the 1 nm Al$_2$O$_3$ contact by a three terminal dV/dI measurement where we inject a current between contacts E2 and E1 and measure voltage between contacts E2 and E3. Figure 1(d) shows typical dV/dI curves that exhibit a cusp-like behavior as a function of DC bias, and junctions with such a shape are found to exhibit little temperature dependence, consistent with tunneling behavior. It is worth noting that as we decrease the barrier thickness, the dV/dI curve transitions from a peak-like shape to a flat shape. Typically, devices with R$_c$A lower than 7 kΩ μm$^2$ exhibit a flat shape dV/dI and an increase of the contact resistance with temperature, suggesting metallic pinhole interface in accordance with a recent TEM study of Al$_2$O$_3$ growth on graphene [33].

Spin transport measurements are carried out using a lock-in detection with an AC injection current of I$_{INJ}$ = 1 μA rms at 11 Hz between contacts E1 and E2 (injector) (see Fig 2a inset). Electron spin density injected at E2 diffuses toward electrode E3 (detector), which generates a non-local voltage V$_{NL}$ measured between E3 and E4. Spin transport is identified by ramping an in-plane magnetic field to achieve parallel and antiparallel alignments of the injector and detector magnetizations. For Hanle spin precession measurements, an out of plane magnetic field is ramped while measuring V$_{NL}$ with electrodes in the parallel and antiparallel magnetization states. For fitting the Hanle data using equation (2) or (4), both the diffusion constant and spin lifetime are fitting parameters. The spin measurements are performed near zero gate voltage (20 out of the 22 samples are measured at V$_G$ = 0 V, while the others have |V$_G$| ≤ 3 V).

**III. Results and Discussion**

We investigate spin transport in graphene spin valves with MgO and $Al_2O_3$ barriers. Figures 2a and 2d show the room temperature nonlocal resistance ($R_{NL} = V_{NL}/I_{INJ}$) as an in-plane magnetic field is swept up and down for MgO and $Al_2O_3$ barriers, respectively. The non-local resistance exhibits a sharp switching as the injector and detector transition from the parallel to antiparallel states, indicating spin transport in graphene. For the MgO device, the size of the nonlocal magnetoresistance ($\Delta R_{NL}$) is ~25 $\Omega$ with the average contact resistance $R_C$ of 5.5 k$\Omega$. For the $Al_2O_3$ device, we measure $\Delta R_{NL}$ of 20 $\Omega$ with average $R_C$ of 58 k$\Omega$.

In order to extract the spin lifetime and spin diffusion length, we perform Hanle spin precession measurements by applying an out-of-plane magnetic field. This causes the spins to precess in-plane as they diffuse between contacts E2 and E3, resulting in the so-called Hanle curves shown in Figure 2(b) and (e) for MgO and $Al_2O_3$, respectively. The red (black) curves correspond to the spin precession with electrodes in the parallel (antiparallel) configuration. The spin lifetime is first extracted using the traditional model by fitting the difference of the parallel and antiparallel Hanle curves (shown in Figure 2(c) and 2(f)) using equation 2, yielding $\tau_{TM}$ = 768 ps for MgO and $\tau_{TM}$ = 685 ps for $Al_2O_3$. Fitting the same data with the spin absorption model using equation 4 yields spin lifetimes of $\tau_{SAM}$ = 954 ps for MgO and $\tau_{SAM}$ = 754 ps for $Al_2O_3$. Because $\tau_{TM}$ includes the effects of both spin relaxation and spin absorption (equation 3), this results in lower effective spin lifetime compared to $\tau_{SAM}$ (equation 5).

In a previous study, Idzuchi *et al* reported an intrinsic spin lifetime in graphene independent of the type of the contact (transparent, pinhole or tunneling) when spin absorption is taken in account, suggesting the absence of the additional contact-induced spin dephasing term.[30] We investigate this further by performing more extensive studies of spin lifetime as a function of contact resistance area product for MgO and $Al_2O_3$ tunnel barriers.

In Figure 3(a) and (d), we present the extracted spin lifetime from experimental Hanle curves using the standard Bloch equation that does not separate out the effect of spin absorption (equation (2)). The spin lifetime extracted from this fitting is an effective lifetime $\tau_{TM}$ that includes the bulk spin lifetime as well as the spin absorption and other contact induced effects following equation (3). Thus the fitted lifetime could be much lower than the bulk spin lifetime $\tau_s$. The effective spin lifetime from this model shows a strong dependence on $R_CA$ (the average resistance-area product of the injector and detector contacts). Specifically, low $R_CA$ corresponds to short $\tau_{TM}$ while large $R_CA$ corresponds to long $\tau_{TM}$. This indicates the importance of contact induced spin relaxation, and the trend is consistent with the expected behavior of $\tau_{TM}$ with $R_CA$ due to spin absorption effect. In order to test this quantitatively, we employ models that separate out the effect explicitly such as the models proposed by van Wees, Aji and Otani[9,29,28].

In Figure 3(b) and (e), we fit the Hanle curves using the Aji's analytical solution of the van Wees model including the spin absorption effect. In this fitting, we use separate values for contact resistances $R_C^1$ and $R_C^2$ of the spin injector and detector although we plot the data based on the average $R_CA$ product. We observe that the extracted value of $\tau_{SAM}$ is larger than $\tau_{TM}$ from the traditional fit for both the $Al_2O_3$ barrier and the MgO. Most strikingly, there is still a strong dependence on the $R_CA$ product of the contacts, contrary to the trend reported in Idzuchi et al.[30] Because the value of $\tau_{SAM}$ does not converge to a universal value independent of $R_CA$, according to equation (5) this implies that other contact induced effects ($\Gamma_{other}$) are important in agreement with *Volmer et al.* [22,23]

In Figure 3(b) we plot the transparent contact data along with the MgO tunnel barrier data for comparison. It is worth noting, that the extracted spin lifetime $\tau_{SAM}$ is very sensitive to the value of the contact resistance of the injector and detector. Therefore, any uncertainty in the contact resistance value will lead to a significant variation in $\tau_{SAM}$. This sensitivity is more prominent in case of transparent devices where the contact resistance measurement includes the Co/graphene junction resistance and the Co lead resistance. Due to the low contact resistance value of the interface, this leads to high uncertainty of the interface resistance and the extracted spin lifetime. In the case of tunnel barrier, the junction resistance is significantly higher than the lead resistance, resulting in a more accurate determination of the spin lifetime. For transparent devices without a tunnel barrier, we observe an increase of $\tau_{SAM}$ by almost two orders of magnitude up to 679 ps as compared to $\tau_{TM}$ = 76 ps for the lowest $R_CA$ data point, consistent with the result in Idzuchi et al.[30] However, as $R_CA$ increases, we observe a decrease in $\tau_{SAM}$ down to 64 ps. The higher values of $R_CA$ for some transparent devices are likely due to resist residue at the interface of Co/graphene which leads to higher contact resistances. This results in a lower spin lifetime as the $R_CA$ increases, mainly due to less correction from the spin absorption effect.

In Figure 3(c), we estimate the role of spin absorption compared to other spin relaxation mechanisms by plotting the $\tau_{TM}/\tau_{SAM}$ ratio as a function of $R_CA$ product for the MgO samples. The discrepancy between $\tau_{TM}$ and $\tau_{SAM}$ is the greatest for transparent contacts, where the $\tau_{TM}/\tau_{SAM}$ ratio is as low as ~10%. As we introduce the MgO barrier, the $\tau_{TM}/\tau_{SAM}$ ratio increases to the ~60 – 90% range (median value of ~80%). To further investigate this effect, we perform the same analysis for the $Al_2O_3$ samples, with the $\tau_{TM}/\tau_{SAM}$ ratio shown in Figure 3(f). Here, the agreement increases into the ~70% - 100% range (median value of ~90%). The relatively high values of $\tau_{TM}/\tau_{SAM}$ for tunneling contacts indicates that there is reasonable agreement between the spin absorption model and traditional model in many cases. The relative importance of spin absorption compared to the overall spin relaxation is given by $\Gamma_{abs}/(\Gamma_{abs} + \Gamma_{other} + \Gamma_{bulk}) = 1 - \tau_{TM}/\tau_{SAM}$, which has a median value of ~20% for MgO and ~10% for $Al_2O_3$. Therefore, we conclude that spin absorption is not the dominant spin relaxation mechanism, in agreement with Maassen *et al.* [17] and Volmer *et al.* [22,23]

Finally, we perform simulations to provide a guideline for reasonable fitting procedures, considering the influence of the contact resistance and channel length on the Hanle curves. To compare the models, we follow a procedure where we generate Hanle curves using the spin absorption model (equation (4)) and fit these curves with the standard Bloch equation (2).[9] In Figure 4, we show a 2D image plot of the ratio $\tau_{TM}/\tau_{SAM}$ as function of $R_c/R_s$ for different values of $L/\lambda$ where $R_c$ is the contact resistance, $R_s$ is the spin resistance of the graphene, L is the channel length and $\lambda$ is the spin diffusion length. The simulations are performed using D=0.01 $m^2/s$ and $\tau_{SAM}$ = 1 ns. While $\tau_{TM}/\tau_{SAM}$ converges to 1 (blue region) at high $R_c/R_s$, we observe a striking decrease in $\tau_{TM}/\tau_{SAM}$ for very low $R_c/R_s$ (red region). The difference between $\tau_{TM}$ and $\tau_{SAM}$ is magnified as we decrease the channel length L between the injector and detector. First, we notice that Rc/Rs and L/λ have a strong influence on the obtained ratio and therefore dictate how well the standard and spin absorption model agree. This can be understood as follows: as we decrease the Rc/Rs ratio, more spins are absorbed into the ferromagnet due to the conductance mismatch. This results in an apparent low spin lifetime when the effect is not taken in account, which will increase the difference between $\tau_{TM}$ and $\tau_{SAM}$. When L/λ becomes large, the $\tau_{TM}/\tau_{SAM}$ converges to 1, largely independent of Rc/Rs, indicating agreement between the fitting procedures (see case L/λ=100). As we increase the channel length, most of the spin relaxation and diffusion occurs in the graphene channel without interference from the contact. Therefore, performing Hanle measurements using long channel lengths and high contact resistance is recommended in order to minimize the spin absorption effect.

To compare with experimental data, we plot our experimental parameters as data points in the 2D graph with open circles for MgO, triangles for $Al_2O_3$ and diamond for transparent. We observe that for the $Al_2O_3$, most of the devices are located in the blue region consistent with the ratios plotted in Figure 3(f). For MgO tunnel barrier, the devices are spread between the blue and red regions. The effect of spin absorption is more pronounced for the transparent devices as indicated by the points located in the red region.

We finally note that this analysis is performed within the context of 1D models. We are aware of some recent work on metal spin valves using 3D models to analyze spin precession data, which shows that 1D models even explicitly including spin absorption may be inadequate to extract reliable spin lifetimes when L is comparable to, or less than λ.[34] These studies also highlight the importance of having long channel lengths to minimize the effects of contacts.

**IV. Conclusions**

We investigate the contact induced spin relaxation in a large set of graphene spin valve devices using two different types of tunnel barrier, MgO and $Al_2O_3$. We observe a strong dependence of $\tau_{SAM}$ with $R_cA$ for both tunnel barriers. Our analysis suggests that spin relaxation rate from other contact induced effect $\Gamma_{other}$ is significant. The spin absorption has a minor effect on the overall spin relaxation except in some cases with transparent contacts. Instead, other sources of contact-induced spin relaxation (e.g. fringe fields, etc.) are more important. Thus, further investigation of the tunnel barrier microstructure using TEM is needed to understand the role of the contact on the spin relaxation mechanism. For tunneling contacts, the $\tau_{TM}/\tau_{SAM}$ ratio has a median value of ~80% for MgO barriers and ~90% for $Al_2O_3$ barriers, indicating that the two models have reasonable agreement in many cases. Nevertheless, the spin absorption models are more accurate and should be applied whenever accurate measurement of the contact resistance can be obtained. Finally, we provide a guideline for estimating the relative importance of spin absorption in determining spin lifetimes through simulations.

**Acknowledgements**

We acknowledge T. Zhu, S. Singh, J. Katoch, P. A. Crowell and G. Stecklein for fruitful discussions and technical assistance. This work was supported by NRI-NSF (DMR-1124601), ONR (N00014-14-1-0350), NSF (DMR-1310661), and C-SPIN, one of the six SRC STARnet Centers, sponsored by MARCO and DARPA.

**References**

1. D. Huertas-Hernando, F. Guinea, and A. Brataas, Phys. Rev. B **74**, 155426 (2006).
2. H. Min, J. E. Hill, N. A. Sinitsyn, B. R. Sahu, L. Kleinman, and A. H. MacDonald, Phys. Rev. B **74**, 165310 (2006).
3. Y. Yao, F. Ye, Y.-L. Qi, S.-C. Zhang, and Z. Fang, Phys. Rev. B **75**, 041401(R) (2007).
4. B. Trauzettel, D. V. Bulaev, D. Loss, and G. Burkard, Nature Physics **3**, 192 (2007).
5. D. Huertas-Hernando, F. Guinea, and A. Brataas, Phys. Rev. Lett. **103**, 146801 (2009).
6. W. Han, R. K. Kawakami, M. Gmitra, and J. Fabian, Nature Nano. **9**, 794 (2014).
7. N. Tombros, C. Jozsa, M. Popinciuc, H. T. Jonkman, and B. J. Van Wees, Nature **448**, 571 (2007).

8. C. Jozsa, T. Maassen, M. Popinciuc, P. J. Zomer, A. Veligura, H. T. Jonkman, and B. J. van Wees, Phys. Rev. B **80**, 241403(R) (2009).

9. M. Popinciuc, C. Jozsa, P. J. Zomer, N. Tombros, A. Veligura, H. T. Jonkman, and B. J. van Wees, Phys. Rev. B **80**, 214427 (2009).

10. M. Shiraishi, M. Ohishi, R. Nouchi, T. Nozaki, T. Shinjo, and Y. Suzuki, Adv. Func. Mater. **19**, 1 (2009).

11. K. Pi, W. Han, K. M. McCreary, A. G. Swartz, Y. Li, and R. K. Kawakami, Phys. Rev. Lett. **104**, 187201 (2010).

12. W. Han, K. Pi, K. M. McCreary, Y. Li, J. J. I. Wong, A. G. Swartz, and R. K. Kawakami, Phys. Rev. Lett. **105**, 167202 (2010).

13. W. Han and R. K. Kawakami, Phys. Rev. Lett. **107**, 047207 (2011).

14. W. Han, K. Pi, W. Bao, K. M. McCreary, Y. Li, W. H. Wang, C. N. Lau, and R. K. Kawakami, Appl. Phys. Lett. **94**, 222109 (2009).

15. T.-Y. Yang, J. Balakrishnan, F. Volmer, A. Avsar, M. Jaiswal, J. Samm, S. R. Ali, A. Pachoud, M. Zeng, M. Popinciuc, G. Guntherodt, B. Beschoten, and B. Ozyilmaz, Phys. Rev. Lett. **107**, 047206 (2011).

16. M. Johnson and R. H. Silsbee, Phys. Rev. Lett. **55**, 1790 (1985).

17. T. Maassen, I. J. Vera-Marun, M. H. D. Guimarães, and B. J. van Wees, Physical Review B **86**, 235408 (2012).

18. S. P. Dash, S. Sharma, J. C. Le Breton, J. Peiro, H. Jaffrès, J. M. George, A. Lemaître, and R. Jansen, Physical Review B **84**, 054410 (2011).

19. S. Garzon, I. Žutić, and R. A. Webb, Physical Review Letters **94**, 176601 (2005).

20. J.-H. Park and H.-J. Lee, Physical Review B **89**, 165417 (2014).

21. B. Li, L. Chen, and X. Pan, Applied Physics Letters **98**, 133111 (2011).

22. F. Volmer, M. Drögeler, E. Maynicke, N. von den Driesch, M. L. Boschen, G. Güntherodt, and B. Beschoten, Physical Review B **88**, 161405 (2013).

23. F. Volmer, M. Drögeler, E. Maynicke, N. von den Driesch, M. L. Boschen, G. Güntherodt, C. Stampfer, and B. Beschoten, Physical Review B **90**, 165403 (2014).

24. A. Varykhalov, D. Marchenko, J. Sánchez-Barriga, M. R. Scholz, B. Verberck, B. Trauzettel, T. O. Wehling, C. Carbone, and O. Rader, Physical Review X **2**, 041017 (2012).

25. A. Varykhalov and O. Rader, Physical Review B **80**, 035437 (2009).

26. T. Abtew, B.-C. Shih, S. Banerjee, and P. Zhang, Nanoscale **5**, 1902 (2013).

27. F. J. Jedema, H. B. Heersche, A. T. Filip, J. J. A. Baselmans, and B. J. van Wees, Nature **416**, 713 (2002).

28. H. Idzuchi, Y. Fukuma, S. Takahashi, S. Maekawa, and Y. Otani, Physical Review B **89**, 081308 (2014).

29. E. Sosenko, H. Wei, and V. Aji, Physical Review B **89**, 245436 (2014).

30. H. Idzuchi, A. Fert, and Y. Otani, Physical Review B **91**, 241407 (2015).

31. B. Dlubak, P. Seneor, A. Anane, C. Barraud, C. Deranlot, D. Deneuve, B. Servet, R. Mattana, F. Petroff, and A. Fert, Appl. Phys. Lett **97**, 092502 (2010).

32. B. Dlubak, M.-B. Martin, C. Deranlot, K. Bouzehouane, S. Fusil, R. Mattana, F. Petroff, A. Anane, P. Seneor, and A. Fert, Appl. Phys. Lett. **101**, 203104 (2012).

33. B. Canto, C. P. Gouvea, B. S. Archanjo, J. E. Schmidt, and D. L. Baptista, Scientific Reports **5**, 14332 (2015).

34. P. A. Crowell, private communication (2015).

**Figure Captions**

**Fig. 1.** (a) AFM image of 1 nm $Al_2O_3$ on graphene, (b) Raman spectroscopy of graphene before (blue) and after (red) deposition of 1 nm $Al_2O_3$ overlayer, (c) Gate dependence of 4-probe resistance of a graphene spin valve, (d) Three terminal differential contact resistance dV/dI of $Al_2O_3$/graphene junction.

**Fig. 2.** (a, d) Non-local MR of a graphene spin valve with MgO and $Al_2O_3$ barriers, respectively, for in-plane magnetic field swept upward (black) and downward (red), (b, e) Hanle spin precession of a graphene spin valve with MgO and $Al_2O_3$ barriers, respectively, for parallel (red) and antiparallel (black) magnetizations of the injector and detector ferromagnets. (c) Fitting of parallel minus antiparallel Hanle curves using the traditional model (equation 2) and spin absorption model (equation 4). Open circles are the data and the black lines are the best fit curves, which overlap for the two models.

**Fig. 3.** (a, d) $\tau_{TM}$ as a function of $R_cA$ for MgO and $Al_2O_3$ barriers, respectively, (b, e) $\tau_{SAM}$ as a function of $R_cA$ for MgO and $Al_2O_3$ barriers, respectively. (c, f) Ratio $\tau_{TM}/\tau_{SAM}$ as a function of $R_cA$ for MgO and $Al_2O_3$ barriers, respectively. The dashed lines are guides to the eye.

**Fig. 4.** 2D image plot of calculated $\tau_{TM}/\tau_{SAM}$ as a function of $L/\lambda$ and $R_C/R_N$. The plot is generated by simulating Hanle curves using the spin absorption model with $\tau_{SAM} = 1$ ns and D = 0.01 m²/s, and fitting with the traditional model to obtain $\tau_{TM}$. The points represent the values of $L/\lambda$ and $R_C/R_N$ (based on spin absorption model) for the experimental Hanle data with circles for MgO barriers, triangles for $Al_2O_3$ barriers, and diamonds for transparent contacts.

Figure 1

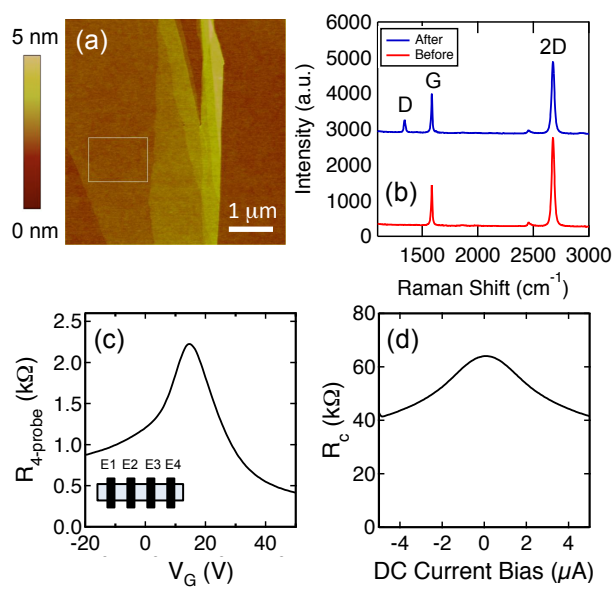

Figure 2

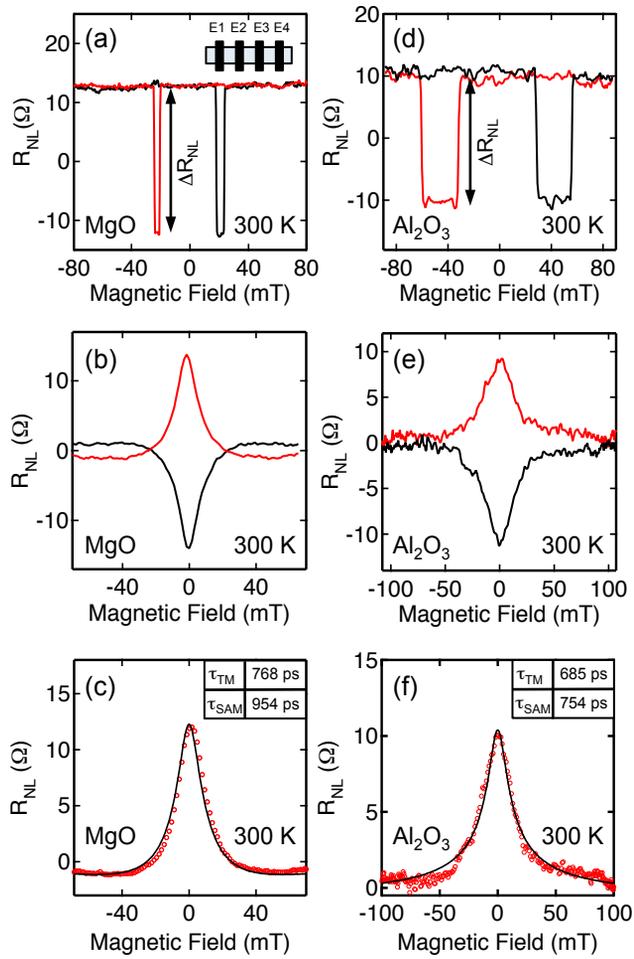

Figure 3

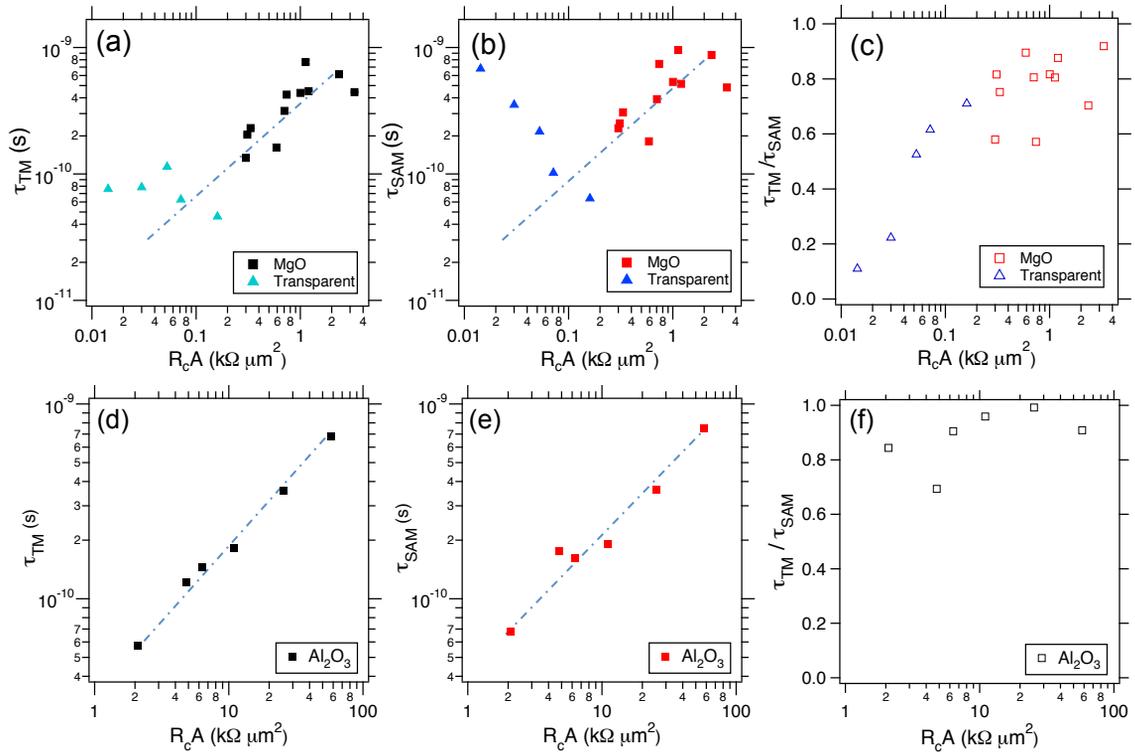

Figure 4

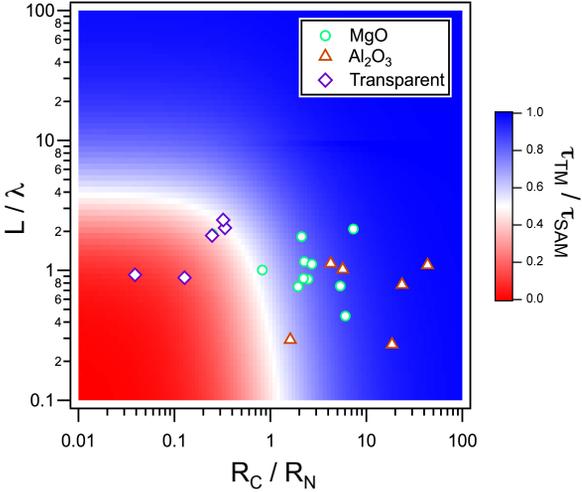